# Compact $U(1) \times U(1)$ Model with Minimal Interspecies Interaction

Ken Yee

Dept. of Physics and Astronomy, L.S.U.

Baton Rouge, Louisiana   70803-4001

email: kyee@rouge.phys.lsu.edu

**Abstract**

We introduce a minimally interacting pure gauge compact $U(1) \times U(1)$ model consistent with abelian projection symmetries. This paradigm, whose interactions are entirely due to compactness, illustrates how compactness can contribute to interspecies interactions. Furthermore, it has a much richer phase structure than obtained by naively tensoring together two compact $U(1)$ copies.

# 1  Compact $U(1) \times U(1)$

While the compact $U(1) \times U(1)$ models suggested in this paper are interesting in their own right as extensions of compact QED [1] to multiple gauge fields, our interest is inspired by the abelian projection [2]. On the lattice the abelian projection of $SU(3)$ gauge configurations [3, 4] yields a $U(1) \times U(1)$ invariant lattice gauge theory which we call "APQCD(abelian projected QCD)." A working hypothesis, supported by recent numerical results in lattice $SU(2)$ and $SU(3)$ [5], is that APQCD captures essential features of QCD confinement. To facilitate progress, it would help to have an analytical paradigm of APQCD.

In numerical studies of APQCD authors typically imagine that one of the $U(1)$ copies is integrated out leaving a single "representative" $U(1)$ species. While there is nothing wrong with this approach, we suggest that there may be some interest if not advantage to thinking about APQCD as a $U(1) \times U(1)$ model rather than a single $U(1)$ species with its partner integrated out. One reason is because the two $U(1)$ gauge fields are correlated [4], that is, they interact. Obviously integrating out a $U(1)$ copy surrenders dynamical information about the interspecies interaction which only a $U(1) \times U(1)$ formulation can give.

Another reason is as follows. For an operator $\mathcal{O}_A^t$, let superscript $t \in \{1, 2\}$ refer to the two $U(1)$ species and subscript $A \in \{l, P, c\}$ indicate whether $\mathcal{O}_A^t$ is a link, plaquette, or cube variable. In addition to gauge and other symmetries APQCD is invariant under

- (A) exchange of species $t = 1$ with $t = 2$;
- (B) charge conjugation $\Theta_l^t \to -\Theta_l^t$ for $t = 1, 2$ simultaneously.

Symmetry $(A)$ arises from the residual permutation gauge invariance preserved by the abelian projection [4], and $(B)$ is due to invariance under hermitian conjugation $U_{x,\mu} \to U_{x,\mu}^{\dagger}$ of the $SU(3)$ links. Restricting ourselves to $1 \times 1$ plaquettes for no other reason than simplicity [6], the two operators which may appear in an action consistent with $(A)$ and $(B)$ are

$$-s_1(\beta, p) \equiv \beta \sum_P \left\{ \cos p\Theta_P^1 + \cos p\Theta_P^2 \right\}, \tag{1}$$

$$-s_2(\gamma_+, \gamma_-, q_+, q_-) = \sum_P \left\{ \frac{\gamma_+}{2} \cos q_+(\Theta_P^1 + \Theta_P^2) + \frac{\gamma_-}{2} \cos q_-(\Theta_P^1 - \Theta_P^2) \right\}. \tag{2}$$

A general APQCD action would be a linear combination of $s_1$ and $s_2$ summed over all integers $p$ and $q_\pm$. Note that while both $s_1$ and $s_2$ are invariant under $2\pi$ shifts of $\Theta_P^t$, $s_2$ is additionally invariant under

$$\Theta_P^t \to \Theta_P^t + (n - (-1)^t m)\pi \qquad n, m \in Z(c_2). \tag{3}$$

In the naive continuum limit $s_1$, which does not induce correlations between species $t = 1$ and $t = 2$, is proportional to $s_0 \equiv \sum_{P,t} \Theta_P^t \Theta_P^t$. In the same limit $s_2$ reduces to

$$-s_2(\gamma_+, \gamma_-, q_+, q_-) \longmapsto \frac{1}{4}(q_+^2 \gamma_+ + q_-^2 \gamma_-)s_0 + H \sum_P \Theta_P^1 \Theta_P^2 \tag{4}$$

and contains the interspecies interaction $\sum_P \Theta_P^1 \Theta_P^2$ if

$$H \equiv \frac{1}{2}(q_+^2 \gamma_+ - q_-^2 \gamma_-) \neq 0. \tag{5}$$

While $H$ is likely nonzero in APQCD we will focus on the $H = 0$ case, which is nontrivial, for simplicity. In Section 2 we show that even if $H$ vanishes interspecies interactions exist due to the compact nature of $s_2$. Thus, in addition to $H \neq 0$ (and possible charged $SU(3)/[U(1)]^2$ matter field) contributions, compactness also contributes to the net APQCD interspecies interaction.

We will refer to $H = 0$ models as "minimal" or "minimally interacting" models. Section 3 argues that the minimal model whose action is

$$S(\beta, p; \gamma, q) \equiv s_1(\beta, p) + s_2(\gamma, \gamma, q, q) \quad p, q \in \mathbf{Z}, \quad \beta, \gamma \in \mathbf{R} \tag{6}$$

has an interesting phase structure. Furthermore, this paradigm illustrates how there may be more than two monopole currents of potential dynamical relevance in a $U(1) \times U(1)$ system. In Toussaint-DeGrand notation [7], the relevant monopoles for $S(\beta, p; \gamma, q)$ are

$$M_c^t = \frac{1}{2\pi} \sum_{P \in c} \left\{ \left(p \Theta_P^t \right) \bmod 2\pi \right\} \quad \forall t \in \{1, 2\}, \tag{7}$$

$$M_c^s = \frac{1}{2\pi} \sum_{P \in c} \left\{ \left(q[\Theta_P^1 - (-1)^s \Theta_P^2] \right) \bmod 2\pi \right\} \quad \forall s \in \{3, 4\} \tag{8}$$

for integers $p$ and $q$. The phase of $S(\beta, p; \gamma, q)$ depends on which combination of $M^1, \cdots, M^4$ is condensed.

Since the underlying action of APQCD is unknown, our results suggest that it may be worthwhile to examine these four monopole operators for a range of $p$ and $q$ in APQCD to see if they are all condensed [8]. Also, we note that if $\gamma \neq 0$, the relationship between the APQCD string tension and APQCD monopoles would be given by a necessarily modified Stack-Wensley formula [9] even should $1^3$ monopoles dominate.

## 2 Interspecies Interaction at $H = 0$

In this Section we focus on the action $s_2$ with parameters $\gamma_\pm = \gamma$ and $q_\pm = 1$ so that interaction coefficient $H = 0$. Adopting lattice differential forms notation [1] define

$$\mathbf{E}(\gamma; \Theta_P) \equiv \sum_{N_P \in Z(c_2)} \exp\left\{-\frac{\gamma}{2} \|\Theta_P - 2\pi N_P\|^2\right\}. \tag{9}$$

In the Villain approximation

$$\exp\{-s_2(\gamma,\gamma,1,1)\} = \mathbf{E}(\frac{\gamma}{2};\Theta_P^1+\Theta_P^2)\mathbf{E}(\frac{\gamma}{2};\Theta_P^1-\Theta_P^2)$$
$$= \{\mathbf{E}(\gamma;\Theta_P^1)\mathbf{E}(\gamma;\Theta_P^2) + \mathbf{E}(\gamma;\Theta^1-\pi)\mathbf{E}(\gamma;\Theta^2-\pi)\} \quad (10)$$

where we have used variables change formula

$$\sum_{n,m\in Z} F(n+m,n-m) = \sum_{N,M\in Z}\{F(2N,2M) + F(2N+1,2M+1)\}. \quad (11)$$

The presence of the second term in (10), which preserves invariance under (3), prevents factorization of $\exp\{-s_2(\gamma,\gamma,1,1)\}$ into independent $\Theta_P^1$ and $\Theta_P^2$ pieces and indicates that $\Theta_P^1$ and $\Theta_P^2$ interact.

To expose the dynamics of this interspecies interaction, consider the Wilson loop expectation value. Upon a BKT transformation [1] in $D=3+1$ dimensions it can be written as

$$\langle W(\mathcal{J}_l)\rangle = \frac{1}{\mathcal{Z}}\exp\{-\frac{1}{2\gamma}\sum_{t=1}^{2}(\mathcal{J}_l^t,\Delta^{-1}\mathcal{J}_l^t)\}\sum_{\substack{\{k_l^s\in Z(c_1)\mid \\ \partial k_l^s=0,\ s=1,2\}}}\exp\{-Y\}, \quad (12)$$

$$Y = \frac{1}{2}\pi^2\gamma\{(k_l^1+k_l^2,\Delta^{-1}(k_l^1+k_l^2)) + (k_l^1-k_l^2,\Delta^{-1}(k_l^1-k_l^2))\} \quad (13)$$
$$+ \pi i\{(^*d(k_l^1+k_l^2),\Delta^{-1}(G_P^1+G_P^2)) + (^*d(k_l^1-k_l^2),\Delta^{-1}(G_P^1-G_P^2))\}.$$

$G_P^{\prime s}$ is related to external electric currents $\mathcal{J}_l^t$ by

$$2\,\partial G_P^{\prime s} = \mathcal{J}_l^1 - (-1)^s \mathcal{J}_l^2 \quad \forall\, s\in\{1,2\}. \quad (14)$$

Hence $\mathcal{J}_l^1$ couples to $\mathcal{J}_l^2$ via the sequence

$$\mathcal{J}_l^1 \leftrightarrow G_P^1 \leftrightarrow \{k_l^1\pm k_l^2\} \leftrightarrow G_P^2 \leftrightarrow \mathcal{J}_l^2, \quad (15)$$

that is, monopole currents $k_l^1$ and $k_l^2$ arising from the periodic nature of $s_2$ mediate interactions between the two $U(1)$ species. Note that using (11) to make variables change $k_l^1\pm k_l^2 \to K_l^\pm$ leads to the BKT-transformed version of (10). One can extend this analysis to $H\neq 0$ models.

# 3  Phases of the Minimal Model

In this Section we derive the phase diagram for the model defined by action $S$ of Eq. (6). Let the Wilson loop, carrying a superposition of electric charges $p$ and $q$, be defined by integer currents

$$\mathcal{J}_l^t \equiv p\mathcal{P}_l^t + q\mathcal{Q}_l^t, \qquad \mathcal{P}_l^t,\ \mathcal{Q}_l^t \in Z(c_1) \tag{16}$$

where $\partial \mathcal{J}_l^t = \partial \mathcal{P}_l^t = \partial \mathcal{Q}_l^t = 0$. For any integer $m$ current $\mathcal{J}_l^t$ is ambiguous under shifts

$$\mathcal{P}_l^t \to \mathcal{P}_l^t + mq, \qquad \mathcal{Q}_l^t \to \mathcal{Q}_l^t - mp. \tag{17}$$

Upon a character expansion the Wilson loop expectation value is

$$\langle W(\mathcal{J}_l) \rangle = \frac{1}{\mathcal{Z}} \sum_{\{\nu_P^s \in Z(c_2) | s=1,2,3,4\}} \int_{\Theta_l^1, \Theta_l^2} \exp\{-X\}, \tag{18}$$

$$X \equiv \sum_{t=1}^{2} \left\{ \frac{1}{2\beta}\left(\nu_P^t, \nu_P^t\right) + \frac{1}{\gamma}\left(\nu_P^{t+2}, \nu_P^{t+2}\right) + i\left(p\partial \nu_P^t + q\partial \overline{\nu}_P^t - \mathcal{J}_l^t, \Theta_l^t\right) \right\} \tag{19}$$

where $\overline{\nu}_P^t \equiv \nu_P^3 - (-1)^t \nu_P^4$. As shown in Appendix A, BKT transformation of (18) yields

$$\langle W(\mathcal{J}_l) \rangle = \frac{1}{\mathcal{Z}} \sum_{\substack{\{j_l^t \in Z(c_1) | \partial j_l^t = 0,\ t=1,2\} \\ \{k_l^s \in Z(c_1) | \partial k_l^s = 0,\ s=1,2,3,4\}}} \exp\{-Y^{\mathrm{diag}} - Y^{\mathrm{int}}\}, \tag{20}$$

$$Y^{\mathrm{diag}} = \sum_{t=1}^{2} \Big\{ 2\pi^2 \beta \left(k_l^t, \Delta^{-1} k_l^t\right) + 4\pi^2 \gamma \left(k_l^{2+t}, \Delta^{-1} k_l^{2+t}\right) \tag{21}$$

$$+\ \frac{1}{2\beta}\left(qj_l^t + \mathcal{P}_P^t, \Delta^{-1}(qj_l^t + \mathcal{P}_P^t)\right) + \frac{1}{2\gamma}\left(pj_l^t - \mathcal{Q}_l^t, \Delta^{-1}(pj_l^t - \mathcal{Q}_l^t)\right) \Big\},$$

$$Y^{\mathrm{int}} = 2\pi i \sum_{s=1}^{4} \left(^*dk_l^s, \Delta^{-1} G_P^s\right). \tag{22}$$

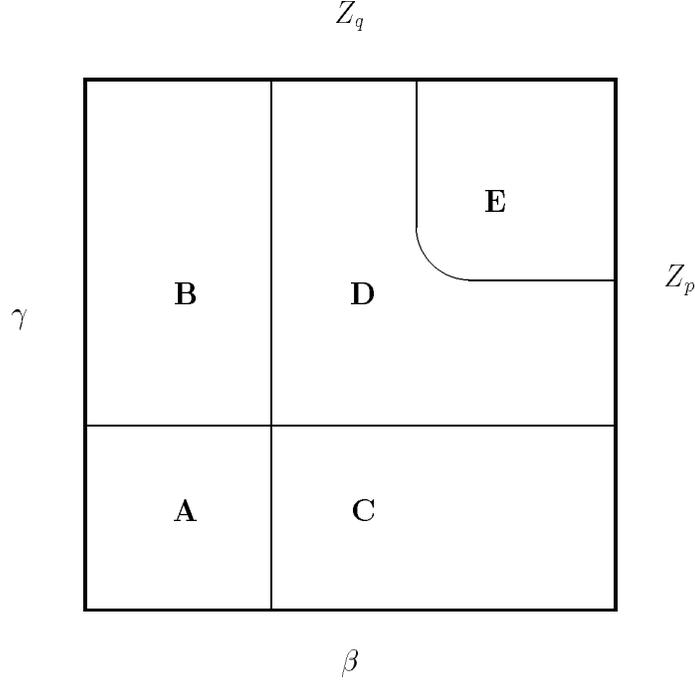

Figure 1: Schematic diagram of the five phases $A - E$ (defined in text) when $5 \leq p < q$ and $\beta \geq 0$ and $\gamma \geq 0$. The $\beta \to \infty$ line is a $Z_p$ model and the $\gamma \to \infty$ line is a $Z_q$ model. This diagram applies to both $U(1)$ species.

Hence the vacuum of this model consists of four fluctuating monopole current loops $k_l^s$ and two electric current loops $j_l^t$. $G_P^s$ is determined by $j_l^t$ and $\mathcal{J}_l^t$ according to (A.1), (A.4) and (A.5). Interspecies interaction—the interaction between $\mathcal{J}_l^1$ and $\mathcal{J}_l^2$—occurs via the sequence

$$\mathcal{J}_l^1 \leftrightarrow j_l^1 \leftrightarrow \{G_P^3, G_P^4\} \leftrightarrow j_l^2 \leftrightarrow \mathcal{J}_l^2. \tag{23}$$

$\langle W(\mathcal{J}_l) \rangle$ in (20) is invariant under current ambiguity (17).

From (20) one can draw by inspection the phase diagram for positive $\beta$ and $\gamma$ as follows. Assume that the gauge fields are rescaled so that integers $p, q$ are either relatively prime or equal. According to the entropy-action balance criterion, loops condense when their path entropy overcomes their Boltzmann weight suppression [1]. By inspection of Eq. (20) $k_l^t$ loops condense when $\beta < \beta_c$ and are suppressed when $\beta > \beta_c$. Similarly $k_l^t$ loops condense when $\gamma < \beta_c$ and are suppressed when $\gamma > \beta_c$. Electric $j_l$ loops condense when $2\pi^2(\frac{q^2}{\beta} + \frac{p^2}{\gamma}) < \beta_c$ and are suppressed when $2\pi^2(\frac{q^2}{\beta} + \frac{p^2}{\gamma}) > \beta_c$. Therefore, as depicted in Figure 1, we predict five phases:

- **A**: Confinement of electrically charged particles whose charges are linear combinations of $p$ and $q$ multiples.

- **B**: Confinement of particles whose electric charges are $p$ multiples; charges which are $q$ multiples experience Coulomb forces.

- **C**: Same as phase **B** with $p \leftrightarrow q$.

- **D**: Nonconfinement phase. $p$ and $q$ charges experience Coulomb forces.

- **E**: Magnetic confinement phase. Electrically charged particles are *not* confined.

On the $\beta \to \infty$ line links are forced to assume $Z_p$ values by $s_1$ and our model reduces to a $Z_p \times Z_p$ model. Analogously, on the $\gamma \to \infty$ line our model reduces to a $Z_q \times Z_q$ model (with variables $\Theta_l^1 \pm \Theta_l^2$). On either the $\beta = 0$ or $\gamma = 0$ lines our model reduces to simpler $U(1) \times U(1)$ models. On these four lines Figure 1 reduces to the phase diagrams [10] respectively of the simple $Z_p$, $Z_q$, $U(1)$ and $U(1)$ models.

Figure 1 is qualitatively the same as the phase diagram of the "mixed" $U(1)$ model with a *single* $U(1)$ species whose action is [11]

$$S^{\text{mixed}} \equiv \beta \cos p\Theta_P + \gamma \cos q\Theta_P. \qquad (24)$$

In other words our minimal $U(1) \times U(1)$ model is similar to *two* copies of the mixed $U(1)$ model. As such we refer the Reader to Refs. [11, 12] for accounts of what happens if $p$ or $q$ is less than 5, in which case some phases merge or disappear. Furthermore, by the symmetry described in Ref. [12], the phase diagram is symmetric under $\beta \to -\beta$, and on the $\beta = 0$ line the phase diagram is symmetric under $\gamma \to -\gamma$. Remarks in [11, 12] concerning whether the phase boundaries are first or second order are superceded by more recent simulations [13] which tend to indicate that these transitions are mostly first order.

## 4 Acknowledgments

I thank D. Haymaker, M. Polikarpov, T. Suzuki, H. Trottier, M. Weinstein, R. Wensley, and R. Woloshyn for stimulating remarks, and TRIUMF and SLAC for the opportunity to present seminars on these results. Related computing work was done at the NERSC, Livermore supercomputer center. The author is supported by DOE grant DE-FG05-91ER40617.

# Appendix A   BKT Transformation

Here we sketch how (20) is obtained from (18). Hodge decomposition yields

$$\nu_P^s \equiv \partial \lambda_c^s + G_P^s, \quad d\lambda_c^s = 0, \quad a_l^s \equiv \partial G_P^s \quad \forall \; s \in \{1,2,3,4\} \qquad (A.1)$$

where $\lambda_c^s \in Z(c_3)$ and $a_l^s \in Z(c_1)$. This decomposition induces a change of summation variables on the RHS of (18) according to

$$\sum_{\{\nu_P^s \in Z(c_2)\}} f(\nu_P^s) \propto \sum_{\substack{\{\lambda_c^s \in Z(c_3) | d\lambda_c^s = 0\} \\ \{a_l^s \in Z(c_1) | \partial a_l^s = 0\}}} f(\partial \lambda_c^s + G_P^{\prime s}). \tag{A.2}$$

Then integration of $\int_{\Theta_l^1, \Theta_l^2}$ yields two constraints

$$pa_l^t + q(a_l^3 - (-1)^t a_l^4) = q\mathcal{Q}_l^t + p\mathcal{P}_l^t \quad \forall\, t \in \{1,2\}. \tag{A.3}$$

We have invoked (16) for $\mathcal{J}_l$. Solutions parametrized by integer currents loops $j_l^t$ to (A.3) are

$$a_l^t = qj_l^t + \mathcal{P}_l^t \quad \forall\, t \in \{1,2\} \tag{A.4}$$

$$2a_l^s = -p(j_l^1 - (-1)^s j_l^2) + \mathcal{Q}_l^1 - (-1)^s \mathcal{Q}_l^2 \quad \forall\, s \in \{3,4\} \tag{A.5}$$

where $j_l^t$ and $\mathcal{Q}_l^t$ must be such that the RHS of (A.5) is even. These solutions allow us to replace the four sums over $\{a_l^s\}$ in (A.2) with two sums over electric current loops $j_l^1$ and $j_l^2$. Finally, Poisson resummation and absorption of irrelevant factors into normalization $\mathcal{Z}$ yields (20).